\documentclass[10pt]{article}
\usepackage{bbm}
\usepackage[final]{graphics}
\usepackage{amsmath,pstricks}
\usepackage{amsfonts,amsbsy}

\font\tenimbf=cmmib10 at 10pt
\font\sevenimbf=cmmib10 at 7pt
\font\fiveimbf=cmmib10 at 5pt
\newfam\imbf
\textfont\imbf=\tenimbf
\scriptfont\imbf=\sevenimbf
\scriptscriptfont\imbf=\fiveimbf

\def\empile#1\over#2{\mathrel{\mathop{\kern 0pt#1}\limits_{#2}}}

\def\bs{\boldsymbol}

\newcommand{\slv}{\raise.15ex\hbox{$/$}\kern-.53em\hbox{$v$}}
\newcommand{\slF}{\raise.15ex\hbox{$/$}\kern-.53em\hbox{$F$}}
\newcommand{\slL}{\raise.15ex\hbox{$/$}\kern-.53em\hbox{$L$}}
\newcommand{\slP}{\raise.15ex\hbox{$/$}\kern-.53em\hbox{$P$}}
\newcommand{\slp}{\raise.15ex\hbox{$/$}\kern-.53em\hbox{$p$}}
\newcommand{\slq}{\raise.15ex\hbox{$/$}\kern-.53em\hbox{$q$}}
\newcommand{\slR}{\raise.15ex\hbox{$/$}\kern-.53em\hbox{$R$}}
\newcommand{\slQ}{\raise.15ex\hbox{$/$}\kern-.53em\hbox{$Q$}}
\newcommand{\slK}{\raise.15ex\hbox{$/$}\kern-.53em\hbox{$K$}}
\newcommand{\slk}{\raise.15ex\hbox{$/$}\kern-.53em\hbox{$k$}}
\newcommand{\slD}{\raise.15ex\hbox{$/$}\kern-.53em\hbox{$D$}}
\newcommand{\slA}{\raise.15ex\hbox{$/$}\kern-.53em\hbox{$A$}}
\newcommand{\slSigma}{\raise.15ex\hbox{$/$}\kern-.53em\hbox{$\Sigma$}}
\newcommand{\slpartial}{\raise.15ex\hbox{$/$}\kern-.53em\hbox{$\partial$}}
\newcommand{\slcalP}{\raise.15ex\hbox{$/$}\kern-.63em\hbox{$\cal P$}}

\def\k{{\boldsymbol k}}

\def\x{{\boldsymbol x}}
\def\y{{\boldsymbol y}}

\def\wk{\omega_{k}}

\def\oeta{\overline{\eta}}

\catcode`\@=11

\newcount\@tempcntc
\def\@citex[#1]#2{\if@filesw\immediate\write\@auxout{\string\citation{#2}}\fi
  \@tempcnta\z@\@tempcntb\m@ne\def\@citea{}\@cite{%
        \@for\@citeb:=#2\do%
    {\@ifundefined{b@\@citeb}%
        {\@citeo\@tempcntb\m@ne\@citea%
                \def\@citea{,\penalty\@m\ }{\bf ?}\@warning%
                {Citation `\@citeb' on page \thepage \space undefined}}%
        {\setbox\z@\hbox{\global\@tempcntc0\csname b@\@citeb\endcsname\relax}
     \ifnum\@tempcntc=\z@ \@citeo\@tempcntb\m@ne%
       \@citea\def\@citea{,\penalty\@m}%
       \hbox{\csname b@\@citeb\endcsname}%
     \else%
      \advance\@tempcntb\@ne%
      \ifnum\@tempcntb=\@tempcntc%
      \else\advance\@tempcntb\m@ne\@citeo%
      \@tempcnta\@tempcntc\@tempcntb\@tempcntc\fi\fi}}\@citeo}{#1}}%

\def\@citeo{\ifnum\@tempcnta>\@tempcntb\else\@citea
  \def\@citea{,\penalty\@m}%
  \ifnum\@tempcnta=\@tempcntb\the\@tempcnta\else
   {\advance\@tempcnta\@ne\ifnum\@tempcnta=\@tempcntb \else
\def\@citea{--}\fi
    \advance\@tempcnta\m@ne\the\@tempcnta\@citea\the\@tempcntb}\fi\fi}

\catcode`\@=12

\begin{document}
\title{\bf Semiclassical thermodynamics\\ of scalar fields}
\author{A. Bessa$^{(1)}$, C.A.A. de Carvalho$^{(1)}$, 
E.S. Fraga$^{(1)}$, F. Gelis$^{(2)}$}
\maketitle
\begin{center}
\begin{enumerate}
\item Instituto de F\'\i sica\\
Universidade Federal do Rio de Janeiro\\
C.P. 68528, Rio de Janeiro, RJ 21941-972, Brazil
\item Theory Division\\
  PH-TH, Case C01600, CERN,\\
  CH-1211 Geneva 23, Switzerland
\end{enumerate}
\end{center}

\begin{abstract}
  We present a systematic semiclassical procedure to compute the
  partition function for scalar field theories at finite
  temperature. The central objects in our scheme are the solutions of
  the classical equations of motion in imaginary time, with spatially
  independent boundary conditions. Field fluctuations -- both field
  deviations around these classical solutions, and fluctuations of the
  boundary value of the fields -- are resummed in a Gaussian
  approximation. In our final expression for the partition function,
  this resummation is reduced to solving certain ordinary differential
  equations. Moreover, we show that it is renormalizable with the
  usual 1-loop counterterms.
\end{abstract}
\vskip 5mm
\begin{flushright}
Preprint CERN-PH-TH/2007-076
\end{flushright}

\section{Introduction}
\label{sec:intro}
Finite-temperature field theory \cite{lebellac} is the natural
framework for the study of phase transitions, and of the thermodynamic
properties of equilibrium states. Applications range from the
investigation of the phase structure of the strong and electroweak
interactions, and the related applications to the early universe, to
the low-energy effective field theories in particle physics and
condensed matter systems.

However, finite-temperature field theories often face a major
difficulty: the plain perturbation expansion
\cite{Arnold:1994ps,Arnold:1994eb,Parwani:1994zz,Kajantie:2001hv} is
ill-defined due to the presence of infrared divergences in the bosonic
sector, and often gives meaningless results.  In the case of hot QCD,
for instance, one can say that the domain of validity of the naive
weak-coupling expansion is the empty set \cite{Braaten:2002wi}.  This
challenge stimulated the development of resummation techniques that
reorganize the perturbative series, and resum certain classes of
diagrams, thereby improving the perturbative expansion (see
\cite{Kraemmer:2003gd,Andersen:2004fp} for recent reviews).  Some of
these techniques amount to using an effective theory in order to 
separate the scales $T$, $gT$, and $g^2T$
\cite{Braaten:1995cm,Braaten:1995ju,Braaten:1995jr,Kajantie:2000iz,Kajantie:2002wa,Kajantie:2003ax}. Others
use modified quasi-particles as the starting point of the perturbative
expansion \cite{Karsch:1997gj}, leading to a significant improvement
of the convergence of the expansion when the mass of these
quasi-particles is properly chosen (one can also mention Refs. 
\cite{Peshier:1994zf,Peshier:1995ty}, where a simple phenomenological
model of massive quasi-particles was successfully used in order to
reproduce the pressure of the quark-gluon plasma obtained in lattice
simulations). In other approaches that aim at maintaining
thermodynamical consistency, one reorganizes the perturbative
expansion of the thermodynamical potential around two-particle
irreducible skeleton diagrams
\cite{Luttinger:1960ua,Blaizot:1999ip,Blaizot:1999ap,Blaizot:2000fc,Blaizot:2003tw,Peshier:1998rz}. Finally,
some of these techniques are based on a systematic use of the Hard
Thermal Loop effective action
\cite{Andersen:1999fw,Andersen:1999sf,Andersen:1999va,Andersen:2000yj,Andersen:2002ey},
i.e., on the assumption that the HTLs provide a good description of the
quasi-particles in the plasma, and of their interactions.

A somewhat different approach, which can also be interpreted as a
resummation of an infinite set of perturbative diagrams, is provided
by the semiclassical approximation \cite{rajaraman,zinn-justin}.
Since the partition function of a given system can be cast in the form
of a path integral whose weight is the exponential of minus the
Euclidean action, an expansion around {\it Euclidean} classical
solutions is quite natural. This program has been carried out in the
case of one-dimensional quantum statistical mechanics for particles in
a single-well potential in \cite{deCarvalho:1998mv}, and also for 
double wells \cite{deCarvalho:2001vk}. From the mere
knowledge of the classical Euclidean solutions of the equation of
motion, the full semiclassical series for the partition function was
constructed\footnote{The equivalent problem in quantum mechanics at
zero temperature was previously studied by DeWitt-Morette
\cite{morette1} for arbitrary potentials, and by Mizrahi
\cite{mizrahi} for the single-well quartic anharmonic oscillator,
using similar techniques. For a more complete list of references on
the semiclassical series in quantum mechanics, see
\cite{deCarvalho:1998mv}}.  Later, these results were generalized to
the case of a particle in a central potential in an arbitrary number
of dimensions \cite{deCarvalho:1999fi}. In both cases, excellent
results were obtained, for instance, for the ground state energy and
the specific heat, in the case of the quartic potential.

In this paper we develop a similar semiclassical procedure to compute
the partition function for thermal scalar field theories with a
single-well potential. We expand around Euclidean classical fields,
whose value on the boundaries of the time interval are taken to be
independent of space. These solutions are assumed to be known to all
orders in the interaction potential (either analytically or
numerically). Then, we incorporate fluctuations around these classical
trajectories, as well as space-dependent fluctuations of the boundary
value of the classical fields. All these fluctuations are kept only in
a Gaussian approximation, although it is in principle possible to go
systematically beyond this approximation. We also provide a
diagrammatic interpretation of our results, connecting our formalism
to the ordinary perturbative expansion, and identifying the classes of
diagrams that are resummed in our approach.  The implementation of the
renormalization procedure in this semiclassical treatment is discussed
in detail at the end.

Since the procedure we propose is infrared finite, we believe it
represents an interesting alternative to other rearrangements of
perturbation theory at finite temperature. In this paper, we present
the general semiclassical framework for an arbitrary potential. We
leave the application to the case of a scalar theory with quartic
self-interactions, and comparisons with other methods to a following
publication \cite{next}.

The paper is organized as follows:

In section \ref{sec:class_fluct}, we discuss the computation of the
partition function, starting from its expression in terms of
a functional integral, within the semiclassical approximation. Although
the main result is, of course, well known, we focus our discussion on
the role played by the boundary conditions in Euclidean time at finite
temperature. We also recall the diagrammatic interpretation of
the integration over quadratic fluctuations around the classical
field.

In section \ref{sec:exp_boundary}, we present a systematic procedure to
incorporate effects from fluctuations of the boundary value of the
field in the computation of the partition function. We explain how one
can perform an expansion in those fluctuations in a consistent way,
provided that one knows the classical solutions for the problem with
constant boundary conditions. We derive formulas that incorporate the
effects of these fluctuations up to quadratic order. These formulas
depend only on the classical field itself, and on a basis of solutions
for the equation of motion for small fluctuations around the
classical field.

In section \ref{sec:int_boundary}, we expand the classical action to
second order in the boundary fluctuations, and discuss
diagrammatically the meaning of this expansion in terms of the
boundary value of the field.  This leads to our final expression for
the partition function in terms of quantities that can be
straightforwardly obtained in explicit form for a given potential once
one knows the classical solution mentioned previously (at least
numerically). This expression, however, still needs to be
renormalized.

The renormalization procedure, which resembles the usual perturbative
procedure, is discussed in section \ref{sec:renorm}. There, we show
how to obtain a finite expression for the partition function through
the introduction of only two counterterms in the action, plus the
subtraction of the zero point energy.

We present our conclusions and outlook in section \ref{sec:conclusions}.
Finally, in the appendix, we illustrate the procedure in the case of
the free theory. As mentioned above, the non-trivial example of the
quartic potential will be addressed in detail in another publication.


\section{Small fluctuations around a classical solution}
\label{sec:class_fluct}
We want to calculate the partition function $Z\equiv {\rm Tr}e^{-\beta
H}$ for a system of interacting scalar fields, making use of a
semiclassical approximation. Our starting point is the expression of
$Z$ in terms of path integrals~:
\begin{equation}
Z=\int[D\varphi(\x)]
\int\limits_{\phi(-\beta/2,\x)=\phi(\beta/2,\x)=\varphi(\x)}[D\phi(\tau,\x)]
\;\;\;\; e^{-S_{_{E}}[\phi]}\; ,
\label{eq:Z1}
\end{equation}
where $S_{_E}[\phi]$ is the Euclidean action of the field:
\begin{equation}\label{euclideanaction1}
S_{_E}[\phi]=
\int\limits_{-\beta/2}^{+\beta/2} d\tau d^3\x
\left[\frac{1}{2}\partial_\mu \phi \partial^\mu\phi 
+\frac{1}{2}m^2 \phi^2
+U(\phi)\right]\; .
\end{equation}
Assume, for the time being, that we know the solution $\phi_c(\tau,\x)$
of the classical equation of motion, that takes the value
$\varphi(\x)$ on the boundaries of the time interval~:
\begin{eqnarray}\label{classiceqforphi0}
  &&(\square_{_E}+ m^2)\phi_c(\tau,\x)+U'\left(\phi_c(\tau,\x)\right)=0
  \; ,\nonumber\\
  && \phi_c(-\beta/2,\x)=\phi_c(\beta/2,\x)=\varphi(\x)\; ,
\label{eq:EOMc}
\end{eqnarray}
where we denote by $\square_{_E}\equiv -(\partial_\tau^2+{\bs\nabla}^2)$
the Euclidean D'Alembertian operator.  

A classical solution is a (local) minimum of $S_E$.  Next, in the
functional integration over $\phi(\tau,\x)$ in eq.~(\ref{eq:Z1}), one
assumes that the integral is dominated by field configurations in the
vicinity of that classical solution, i.e., by small fluctuations around 
this classical solution. In order to evaluate the integral in this
approximation, one writes~:
\begin{equation}
\phi(\tau,\x)\equiv\phi_c(\tau,\x)+\eta(\tau,\x)\; ,
\end{equation}
and expands the Euclidean action to second order in the fluctuation
$\eta(\tau,\x)$~:
\begin{eqnarray}\label{Se[fi]up2sndorderinN}
&&S_{_E}[\phi]=
S_{_E}[\phi_c]\nonumber\\
&&\qquad+
\frac{1}{2}
\int (d^4x_1)_{_E} (d^4x_2)_{_E}
\;\left.
\frac{\delta^2 S_{_E}[\phi]}{\delta \phi(x_1)\delta\phi(x_2)}
\right|_{\phi=\phi_c} 
\eta(x_1)\eta(x_2)
+{\cal O}(\eta^4)\; .
\end{eqnarray}
In this equation, we have used the shorthands $x\equiv(\tau,\x)$ and $\int
(d^4x)_{_E}\equiv \int_{-\beta/2}^{\beta/2} d\tau\int d^3\x$.  For the
sake of brevity, let us also introduce the following notation~:
\begin{equation}
\eta^T A[\phi_c] \eta\equiv
\int (d^4x_1)_{_E} (d^4x_2)_{_E}
\;\left.
\frac{\delta^2 S_{_E}[\phi]}{\delta \phi(x_1)\delta\phi(x_2)}
\right|_{\phi=\phi_c} 
\eta(x_1)\eta(x_2)
\; ,
\end{equation}
where $A[\phi_c]$ is a symmetric ``matrix'' that depends on the
classical solution $\phi_c$ (with continuous indices spanning
$[-\beta/2,\beta/2]\times {\mathbbm R}^3$).

The Gaussian functional integration over the fluctuation $\eta$ must
be performed with the constraint that the fluctuation $\eta(\tau,\x)$
vanishes at the time boundaries,
\begin{equation}\label{bcforeta}
\forall\x\;,\quad \eta(-\beta/2,\x)=\eta(\beta/2,\x)=0\; ,
\end{equation}
because the classical background field already saturates the boundary
conditions.  Let us therefore call $A^*[\phi_c]$ the restriction of
the operator $A[\phi_c]$ to the subspace of fluctuations $\eta$ that
obey these boundary conditions.  We can write~:
\begin{equation}
Z\approx \int[D\varphi(\x)] \;
e^{-S_{_E}[\phi_c]}
\;
\left[
{\rm det}\,\left(
A^*[\phi_c]
\right)
\right]^{-1/2}\; .
\label{eq:Z2}
\end{equation}

\begin{figure}[htbp]
\begin{center}
\resizebox*{!}{3cm}{\includegraphics{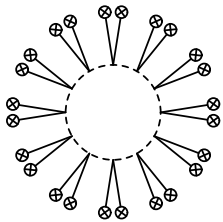}}
\end{center}
\caption{\label{fig:log_Z_phi}Typical 1-loop diagram included in the
integration over fluctuations around the classical solution in the
Gaussian approximation. The lines terminated by a cross denote the
classical solution with a fixed boundary condition $\varphi(\x)$. The
dashed line can be seen as the propagator around the classical
field, for a fluctuation that vanishes at the time boundaries.}
\end{figure}

In order to compute the semiclassical calculation of $Z$, one must
now integrate over the boundary value of the field,
$\varphi(x)$. However, before we pursue this calculation, it is useful
to recall the nature of the diagrams that are contained in the square
root of the functional determinant. It is well known that the Gaussian
integration over fluctuations above a given background field amounts
to calculating the one-loop correction to the effective
action. However, for this correspondence to be valid, one must
integrate over {\sl all the periodic fields} $\eta(x)$. In our case,
the Gaussian integration involves only fields $\eta$ that vanish on
the time boundaries (see eq.~(\ref{bcforeta})), i.e., only a subset of
all the periodic fields. Therefore, the quantity $\big[{\rm
det}\,\left( A^*[\phi_c] \right)\big]^{-1/2}$ is a part of the
one-loop effective action, but does not contain all the terms that
would normally enter in the effective action at this
order\footnote{This distinction can also be seen by studying the
eigenfunctions of the operator $A[\phi_c]$, on the space of periodic
functions and on the space of functions that vanish at
$\tau=\pm\beta/2$ respectively.}.  With this caveat in mind, a typical
diagram included in this quantity is displayed in figure
\ref{fig:log_Z_phi}, in the case of a field theory with a quartic
coupling. It is important to remember that the propagator represented
by the dashed line differs from the complete time ordered propagator,
because it corresponds to a subset of all the periodic modes.

\begin{figure}[htbp]
\begin{center}
\resizebox*{!}{5cm}{\rotatebox{90}{\includegraphics{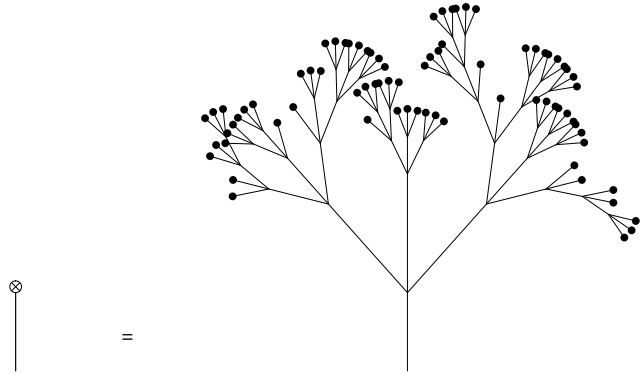}}}
\end{center}
\caption{\label{fig:classfield}Diagrammatic expansion of the classical
field in terms of the boundary value of the field (black dots).}
\end{figure}

Both the classical action and the determinant in eq.~(\ref{eq:Z2})
depend on the field $\varphi(\x)$ on the boundary, through the
dependence of the classical solution $\phi_c$ on the boundary
conditions in eq.~(\ref{eq:EOMc}). In fact, the classical solution
$\phi_c$ can be represented diagrammatically as the sum of all the
tree diagrams terminated by the boundary field $\varphi(\x)$. The
easiest way to see this is to write Green's formula for the solution
of eq.~(\ref{eq:EOMc}). Let us first introduce a Green's function of
the operator $\square_{_E}+m^2$~:
\begin{equation}
\left[\partial_{\tau^\prime}^2+{\bs\nabla}_{\x^\prime}^2
-m^2\right]
G^0(\tau,\x;\tau^\prime,\x^\prime)
=\delta(\tau-\tau^\prime)\delta(\x-\x^\prime)\; .
\label{eq:free_propagator}
\end{equation}
This Green's function is not unique, but we can postpone its choice for
later. Let us multiply this equation by the classical field
$\phi_c(\tau^\prime,\x^\prime)$, and integrate over $\tau^\prime$ and
$\x^\prime$. This gives~:
\begin{equation}
\phi_c(\tau,\x)=
\int_{-\beta/2}^{\beta/2}d\tau^\prime\int d^3\x^\prime\;
\phi_c(\tau^\prime,\x^\prime)
\left[\partial_{\tau^\prime}^2+{\bs\nabla}_{\x^\prime}^2
-m^2\right]G^0(\tau,\x;\tau^\prime,\x^\prime)\; .
\end{equation}
Now, multiply the equation of motion for
$\phi_c(\tau^\prime,\x^\prime)$ by the Green's function
$G^0(\tau,\x;\tau^\prime,\x^\prime)$, integrate over $\tau^\prime$,
and subtract the resulting equation from the previous one. This leads
to~:
\begin{eqnarray}
&&
\phi_c(\tau,\x)
=
\int_{-\beta/2}^{\beta/2}d\tau^\prime\int d^3\x^\prime\;
G^0(\tau,\x;\tau^\prime,\x^\prime)\;
U^\prime(\phi_c(\tau^\prime,\x^\prime))
\nonumber\\
&&
\quad
+\int_{-\beta/2}^{\beta/2}d\tau^\prime\int d^3\x^\prime\;
G^0(\tau,\x;\tau^\prime,\x^\prime)
\Big[
\stackrel{\rightarrow}{\partial_{\tau^\prime}^2}
+
\stackrel{\rightarrow}{{\bs\nabla}_{\x^\prime}^2}
-
\stackrel{\leftarrow}{\partial_{\tau^\prime}^2}
-
\stackrel{\leftarrow}{{\bs\nabla}_{\x^\prime}^2}
\Big]
\phi_c(\tau^\prime,\x^\prime)
\; ,
\nonumber\\
&&
\label{eq:green}
\end{eqnarray}
where the arrows on the differential operators on the second line
indicate on which side they act. The second line can be rewritten as
a boundary term, by noting that~:
\begin{equation}
A\Big[\stackrel{\rightarrow}{\partial_\mu^2}
-
\stackrel{\leftarrow}{\partial_\mu^2}\Big]B
=
\partial^\mu \left\{
A\Big[\stackrel{\rightarrow}{\partial_\mu}
-
\stackrel{\leftarrow}{\partial_\mu}\Big]B
\right\}\; .
\end{equation}
In eq.~(\ref{eq:green}), the boundary in the spatial directions does
not contribute to the classical field at the point $\x$ because the
free propagator decreases fast enough when the spatial separation
increases. Thus, we are left with only a contribution from the
boundaries in time. At this point, since the boundary conditions for
$\phi_c$ consist in specifying the value of the field at
$\tau^\prime=\pm\beta/2$, while its first time derivative is not
constrained, it is very natural to choose a Green's function $G^0$ that
obeys the following conditions\footnote{It is in general always
possible to impose two conditions on a Green's function of
$\square_{_E}+m^2$, because the zero modes of this operator form a
linear space of dimension 2. The conditions of
eq.~(\ref{eq:bound_prop}) are explicitly realized by~:
\begin{eqnarray}
&&
G^0(\tau,\x;\tau^\prime,\x^\prime)
=
\int\frac{d^3\k}{(2\pi)^2}\; e^{i\k\cdot(\x-\x^\prime)}\;
\left\{
\theta(\tau-\tau^\prime)
\frac{\sinh(\omega_\k(\tau-\frac{\beta}{2}))\sinh(\omega_k(\tau^\prime+\frac{\beta}{2}))}
{\omega_\k\,\sinh(\omega_\k\beta)}
\right.
\nonumber\\
&&\qquad\qquad\qquad\qquad\qquad\qquad\qquad\qquad
+
\left.
\theta(\tau^\prime-\tau)
\frac{\sinh(\omega_\k(\tau^\prime-\frac{\beta}{2}))\sinh(\omega_k(\tau+\frac{\beta}{2}))}
{\omega_\k\,\sinh(\omega_\k\beta)}
\right\}\; ,\nonumber
\end{eqnarray}
where we denote $\omega_\k\equiv \sqrt{\k^2+m^2}$.
}~:
\begin{equation}
G^0(\tau,\x;-\beta/2,\x^\prime)
=
G^0(\tau,\x;+\beta/2,\x^\prime)
=0\; .
\label{eq:bound_prop}
\end{equation}
With this choice of the propagator, we obtain the following formula
for $\phi_c(\tau,\x)$~:
\begin{eqnarray}
&&
\phi_c(\tau,\x)
=
\int_{-\beta/2}^{\beta/2}d\tau^\prime\int d^3\x^\prime\;
G^0(\tau,\x;\tau^\prime,\x^\prime)\;
U^\prime(\phi_c(\tau^\prime,\x^\prime))
\nonumber\\
&&
\qquad\qquad\qquad
-\int d^3\x^\prime\;
\varphi(\x^\prime)\;
\Big[
\partial_{\tau^\prime}G^0(\tau,\x;\tau^\prime,\x^\prime)
\Big]_{\tau^\prime=-\beta/2}^{\tau^\prime=+\beta/2}
\; ,
\nonumber\\
&&
\label{eq:green1}
\end{eqnarray}
This formula tells us how the classical solution $\phi_c$ depends on
the boundary value $\varphi(\x)$. If the first term in the right hand
side -- involving the derivative $U^\prime$ of the potential -- were
not there, then the relationship between $\phi_c$ and the boundary
value $\varphi$ would be linear. This only happens in a free
theory. When there are interactions, one can solve
eq.~(\ref{eq:green1}) iteratively in powers of $U^\prime$. This
expansion can be represented diagrammatically by the sum of the tree
diagrams whose ``leaves'' are made of the boundary field
$\varphi(\x)$. An example of such a diagram is illustrated in 
figure \ref{fig:classfield}, in the case of a $\phi^4$ interaction of the fields.
Notice that, when the boundary field is small\footnote{By this, we mean
that the interaction term is smaller than the kinetic term in the
action. This condition depends on the particular momentum modes one is
interested in.}, this sum of trees can be approximated by the
zeroth order in the expansion in powers of $U^\prime$, which is
independent of the interactions. On the other hand, for large values
of $\varphi$, it is important to keep the full sum of tree diagrams
that are summed in $\phi_c$, because all the terms in the expansion
can be equally important. Therefore, we already see an important
feature of our approximation scheme: although the quantum fluctuations
are only included at the 1-loop level, it treats the boundary field to
all orders, allowing a correct treatment even for non-perturbatively
large values of $\varphi(\x)$.

\section{Expansion in fluctuations of the boundary}
\label{sec:exp_boundary}
\subsection{Preliminary discussion}
The integration over boundary configurations that remains to be
performed in eq.~(\ref{eq:Z2}) makes the semiclassical approximation
for $Z$ rather involved; first, we must solve the partial differential
equation~(\ref{eq:EOMc}) for an arbitrary $\varphi(\x)$, and this will
not be feasible analytically in general. Even numerically, this is a
very complicated task. Besides that, the only functional integral over
$\varphi$ that one is able to perform analytically is a Gaussian
integral. In order to circumvent these problems, we are forced to
perform some further approximations.

One can see readily in eqs.~\eqref{eq:EOMc} that the classical
equation of motion reduces to an ordinary differential equation in the
case where the field $\varphi(\x)$ on the boundary is a constant
$\varphi_0$. In this case, the classical solution $\phi_x(\tau,\x)$
becomes a function $\phi_0(\tau)$ of the time only~:
\begin{eqnarray}
&& (-\partial_\tau^2 + m^2)\phi_0(\tau)+U'\left(\phi_0(\tau)\right)=0
\; ,\nonumber\\
&& \phi_0(-\beta/2)=\phi_0(\beta/2)=\varphi_0\; .
\label{eq:EOMcphi=cons}
\end{eqnarray}
Such a simplification of the classical equation of motion would make
the problem much more tractable by analytical or numerical methods.
This remark suggests that we decompose the boundary field
$\varphi(\x)$ into a constant part $\varphi_0$, and a fluctuation
$\xi(\x)$~:
\begin{equation}
\varphi(\x)=\varphi_0+\xi(\x)\; .
\end{equation}
The solution of the classical equation of motion can therefore be
expanded in a similar manner~:
\begin{equation}
\phi_c(\tau,\x)=\phi_0(\tau)+\phi_1(\tau,\x)+\phi_2(\tau,\x)+\cdots\; ,
\label{eq:phic-exp}
\end{equation}
where $\phi_n$ is of order $n$ in $\xi$ (there are terms of
arbitrarily high order in $\xi$ if the equation of motion is non
linear).  Having done this, we can rewrite the path integral over
$\varphi(\x)$ in eq.~(\ref{eq:Z2}) as follows~:
\begin{equation}
Z\approx \int\limits_{-\infty}^{+\infty}d\varphi_0
\int\limits_{\big<\xi(\x)\big>=0} [D\xi(\x)]\;
e^{-S_{_E}[\phi_c]} \;
\left[
{\rm det}\,\left(
A^*[\phi_c]
\right)
\right]^{-1/2}\; .
\label{eq:Z3}
\end{equation}
Notice that the integration over $\xi(\x)$ must be performed with the
constraint that
\begin{equation}\label{Sxid3x=0}
\big<\xi(\x)\big>\equiv\int d^3\x\; \xi(\x)=0\; ,
\end{equation}
since the ``uniform component'' (i.e., the zero mode) of the boundary
condition is already included in $\varphi_0$. An unrestricted
integration of $\xi(\x)$ would therefore overcount the contribution of
this zero mode.

In the following, we are going to assume that the first term
$\phi_0(\tau)$ can be determined with an arbitrary accuracy -- it can
be determined analytically in certain cases, while in general it is
obtained by solving numerically an ordinary differential
equation. Moreover, the dependence on $\phi_0$ will always be treated
exactly. Only the terms that are of higher order in the fluctuation
$\xi$ of the boundary field will be treated in some approximate way.
Doing this allows us to preserve the benefits of treating correctly
the interaction term when the boundary field is large, since only the
fluctuations of the boundary field are assumed to be perturbative.

A natural approximation to obtain the dependence on $\xi$ is to do a
Gaussian approximation around $\xi=0$.  As we shall see
shortly, in order to find the classical action $S_{_E}[\phi_c]$ at
order two in the fluctuation $\xi(\x)$ of the boundary, it is enough
to obtain the classical solution $\phi_c$ at order one in
$\xi(\x)$. 

Moreover, to be consistent with the Gaussian approximation for
$S_{_E}[\phi_c]$, we only need to evaluate the determinant at lowest
order in $\xi(\x)$, i.e., at order zero. Indeed, the Gaussian integration
over the fluctuations $\xi$ corresponds to a one-loop correction in
the background $\phi_0$. However, as we have seen in the previous
section, the functional determinant in eq.~(\ref{eq:Z3}) is already a
one-loop correction. Therefore, keeping the $\xi$ dependence in this
determinant would give higher loop corrections when we integrate over
$\xi$, but only a certain subset of all the 2-loop corrections would
be included. Doing so is not forbidden by any fundamental principle,
but it would arguably make the calculation more complicated; and
moreover this would alter the renormalization of the final
result. Indeed, as we shall see later, by not expanding in $\xi$ the
functional determinant in eq.~(\ref{eq:Z3}), we will eventually obtain
an expression whose ultraviolet divergences are precisely those of the
one-loop effective action. For these reasons, we are going to evaluate
\begin{equation}
Z\approx \int\limits_{-\infty}^{+\infty}d\varphi_0
\;
\left[
{\rm det}\,\left(
A^*[\phi_0]
\right)
\right]^{-1/2}
\int\limits_{\big<\xi(\x)\big>=0} \;
[D\xi(\x)]\;
e^{-S_{_E}[\phi_c]} \; .
\label{eq:Z4}
\end{equation}

\subsection{Correction to $\phi_c$ due to boundary fluctuations}

The next step is to find the correction $\phi_1(\tau,\x)$ to the
classical solution $\phi_c$. In order to find the equation obeyed by
$\phi_1$, simply replace $\phi_c$ by $\phi_0+\phi_1$ in
eq.~(\ref{eq:EOMc}). By dropping all the terms that are of order
higher than unity in $\phi_1$ (since they are at least of order two in
$\xi$), and using the equation obeyed by $\phi_0$, we obtain the
following (linearized) equation for $\phi_1$~:
\begin{equation} 
\Big[(\square_{_E}+m^2)+U''\left(\phi_0(\tau)\right)\Big]\phi_1=0\; ,
\label{eq:EOM1}
\end{equation}
with the boundary condition:
\begin{equation}
\phi_1(-\beta/2,\x)=\phi_1(\beta/2,\x)=\xi(\x)\; .
\end{equation}
In the following, we also need the Green's formula for the variation
$\phi_1$ of the classical field. The derivation is very similar to the
derivation of eq.~(\ref{eq:green1}), and we shall not reproduce it 
here. The main difference compared to eq.~(\ref{eq:green1})
is that we need a Green's function for the operator
$\left(\square_{_E}+m^2+U''\left(\phi_0(\tau)\right)\right)$,
\begin{equation}
\left[\partial_{\tau^\prime}^2+{\bs\nabla}_{\x^\prime}^2
-m^2-U''\left(\phi_0(\tau^\prime)\right)\right]
G(\tau,\x;\tau^\prime,\x^\prime)
=\delta(\tau-\tau^\prime)\delta(\x-\x^\prime)\; ,
\label{eq:propagator}
\end{equation}
instead of the free propagator $G^0$ that we have introduced
earlier. Again, this propagator must obey the boundary condition
\begin{equation}
G(\tau,\x;-\beta/2,\x^\prime)=G(\tau,\x;\beta/2,\x^\prime)=0\; .
\end{equation}
In terms of the fluctuation $\xi$ and of the propagator, the first
order correction to the classical solution reads:
\begin{equation}
\phi_1(\tau,\x)
=
\int d^3\x^\prime\;
\xi(\x^\prime)\left[
\partial_{\tau^\prime}
G(\tau,\x;\tau^\prime,\x^\prime)\right]_{\tau^\prime=-\beta/2}^{\tau^\prime=+\beta/2}\; .
\end{equation}
Notice that, since the background field $\phi_0$ does not depend on
space, the propagator $G$ depends only on the difference
$\x-\x^\prime$. Thus, we can get rid of the spatial convolution by
going to Fourier space:
\begin{equation}
\phi_1(\tau,\k)
=
\xi(\k)
\left[\partial_{\tau^\prime}G(\tau,\tau^\prime,\k)\right]_{\tau^\prime=-\beta/2}^{\tau^\prime=+\beta/2}\; ,
\label{eq:phi1}
\end{equation}
where the  propagator in Fourier space is defined by
\begin{equation}
\left[\partial_{\tau^\prime}^2-(\k^2+m^2)
-U''\left(\phi_0(\tau^\prime)\right)\right]G(\tau,\tau^\prime,\k)
=\delta(\tau-\tau^\prime)\; ,
\label{eq:prop1}
\end{equation}
and
\begin{equation}
G(\tau,-\beta/2,\k)=G(\tau,\beta/2,\k)=0\; .
\label{eq:prop2}
\end{equation}

\subsection{Propagator in the background $\phi_0$}
It is fairly easy to determine the propagator $G$ that obeys
eqs.~(\ref{eq:prop1}) and (\ref{eq:prop2}) in terms of two linearly
independent solutions of the homogeneous linear differential equation:
\begin{equation}
\left[\partial_{\tau}^2-(m^2+\k^2)
-U''\left(\phi_{0}(\tau)\right)\right]
\eta(\tau,\k^2)=0\; .
\label{eq:ode}
\end{equation}
Let $\eta_{a}(\tau,\k)$ and $\eta_{b}(\tau,\k)$ be two such
independent solutions\footnote{When $\k=0$, it is
straightforward to verify that~:
\begin{eqnarray}
&&\eta_a(\tau;0)=\dot{\phi}_0(\tau)\; ,
\nonumber\\
&&\eta_b(\tau;0)=\dot{\phi}_0(\tau)
\int\limits_0^\tau \frac{d\tau^\prime}{{\dot{\phi}}_0^2(\tau^\prime)}\; .
\label{eq:eta-ab}
\end{eqnarray}
obey \eqref{eq:ode}. However, this construction fails when $\k \neq
0$.} of \eqref{eq:ode}. In order to construct from $\eta_{a,b}$ a
solution of eqs.~(\ref{eq:prop1}) and (\ref{eq:prop2}), let us first
introduce the following object:
\begin{equation}
\Omega(\tau,\tau^\prime,\k^2)\equiv
\eta_a(\tau,\k^2)\eta_b(\tau^\prime,\k^2)
-
\eta_b(\tau,\k^2)\eta_a(\tau^\prime,\k^2)\; .
\label{eq:Omega}
\end{equation}
It is trivial to check that $\Omega(\tau,\tau^\prime,\k^2)$ satisfies
eq.~(\ref{eq:ode}), both with respect to the variable $\tau$ and to the
variable $\tau^\prime$. 
Let us then consider the following quantity:
\begin{eqnarray}
&& H(\tau,\tau^\prime,\k^2)\equiv
\frac{\Omega(\beta/2,\tau,\k^2)\Omega(\tau^\prime,-\beta/2,\k^2)}{\Omega(\beta/2,-\beta/2,\k^2)} 
\qquad {\rm if\ } \tau>\tau^\prime\; ,\nonumber\\
&& H(\tau,\tau^\prime,\k^2)\equiv
\frac{\Omega(\beta/2,\tau^\prime,\k^2)\Omega(\tau,-\beta/2,\k^2)}{\Omega(\beta/2,-\beta/2,\k^2)} 
\qquad {\rm if\ } \tau<\tau^\prime\; .
\end{eqnarray}
This quantity obeys eq.~(\ref{eq:prop1}) if
$\tau\not=\tau^\prime$. Moreover, although $H(\tau,\tau^\prime,\k^2)$
is continuous at $\tau=\tau^\prime$, its first time derivative is not,
and one has:
\begin{eqnarray}
&&
\lim_{\epsilon\to 0^+}
\left(
\left.
\partial_{\tau^\prime}H(\tau,\tau^\prime,\k^2)
\right|_{\tau^\prime=\tau+\epsilon}
-
\left.
\partial_{\tau^\prime}H(\tau,\tau^\prime,\k^2)
\right|_{\tau^\prime=\tau-\epsilon}
\right)
\nonumber\\
&&\qquad
=\eta_a(\tau,\k^2)\dot{\eta}_b(\tau,\k^2)
-\dot{\eta}_a(\tau,\k^2)\eta_b(\tau,\k^2)\equiv -W\; .
\end{eqnarray}
as can be checked by an explicit calculation. The right hand side of
the previous equation is nothing but the Wronskian $W$ of the pair of
solutions $\eta_{a,b}$ and is independent of $\tau$ in the case of
eq.~(\ref{eq:ode}). Let us denote by $W$ the value of the Wronskian for
the pair of solutions $\eta_{a,b}$. The discontinuity of
$\partial_{\tau^\prime}H(\tau,\tau^\prime,\k^2)$ across
$\tau^\prime=\tau$ is therefore equal to $W$, which means that the
second time derivative indeed contains a term
$W\,\delta(\tau-\tau^\prime)$.  Finally, from the obvious property
\begin{equation}
\Omega(\tau,\tau,\k^2)=0\; ,
\end{equation}
one easily sees that $H(\tau,\tau^\prime,\k^2)$ satisfies the boundary
condition of eq.~(\ref{eq:prop2}). Therefore,
$W^{-1}\,H(\tau,\tau^\prime,\k^2)$ is the propagator we are looking
for:
\begin{equation}
G(\tau,\tau^\prime,\k)=
-
\frac{\Omega(\beta/2,{\rm max}(\tau,\tau^\prime),\k^2)
\Omega({\rm min}(\tau,\tau^\prime),-\beta/2,\k^2)}
{W\;\Omega(\beta/2,-\beta/2,\k^2)}\;.
\label{eq:G-1}
\end{equation}
In general, the solutions $\eta_{a,b}$ will not be found analytically
for a non-zero $\k$, and will have to be found numerically.

\subsection{Calculation of the functional determinant}
\label{sec:det-Bstar}
As we have already explained, we need to calculate the determinant
that appears in eq.~(\ref{eq:Z3}), $\rm{ det}\,(A^*[\phi_c])$, to
order zero in the fluctuation $\xi(\x)$ of the boundary, i.e., $\rm{
  det}\,(A^*[\phi_0])$. Integrating by parts the kinetic term, the
Euclidean action can be rewritten as~:
\begin{equation}
S_{_E}[\phi_0+\eta]\approx S_{_E}[\phi_0]
+\int (d^4x)_{_E}\;
\left[\frac{1}{2}\eta \square_{_E} \eta 
+\frac{1}{2}m^2 \eta^2
+U^{\prime\prime}(\phi_0)\eta^2\right]\; .
\end{equation}
Notice that the integration by parts does not introduce any boundary
term here, thanks to the boundary condition obeyed by $\eta$ (see
eq.~(\ref{bcforeta})).  Therefore, we have for the operator $A^*$ the
following expression~:
\begin{eqnarray}\label{usual2ndderivative}
A^*[\phi_0]_{\tau,\x;\tau^\prime,\y}&\equiv& 
\left.
\frac{\delta^2 S_{_E}}{\delta\phi(\tau,\x)\delta\phi(\tau^\prime,\y)}
\right|_{\phi=\phi_0}
\nonumber\\
&=&
\delta(\tau-\tau^\prime)\delta(\x-\y)
\left[
\square_{_E}+m^2+U''\left(\phi_0(\tau)\right)
\right]\; .
\end{eqnarray}
Notice that here we have already written $\phi_0$ explicitly as a
field that depends only on time (because we are calculating the
determinant only at order zero in the fluctuations of the
boundary). Thus, we can perform a Fourier transform with respect to
space, and use $\k$ instead of $\x$.  The eigenvalues $g_i$ and
eigenfunctions $\eta_i$ of the operator $A^*[\phi_0]$ are functions
$\eta(\tau,\x)$ that obey the following system of
equations\footnote{For many more informations about properties of
Hill's equations and their solutions, the reader may consult
\cite{hills-equation}.}~:
\begin{eqnarray}
&&\Big[\partial_\tau^2-(m^2+\k^2)
-U''\left(\phi_0(\tau)\right)\Big]\eta_i(\tau,\k^2)
= g_i \eta_i(\tau,\k^2)\; ,
\nonumber\\
&&\forall \x\;,\quad \eta_i(-\beta/2,\k^2)=\eta_i(\beta/2,\k^2)=0\; .
\label{eq:eigen}
\end{eqnarray}
This equation is of the same type as eq.~(\ref{eq:ode}), the only
difference being that $\k^2$ is now replaced by $\k^2+g_i$. Therefore,
it has two independent solutions that are given by
$\eta_a(\tau,\k^2+g_i)$ and $\eta_b(\tau,\k^2+g_i)$, and its general
solution can be written as:
\begin{equation}
\eta_i(\tau,\k^2)=C_a \eta_a(\tau,\k^2+g_i)+C_b \eta_b(\tau,\k^2+g_i)\; ,
\end{equation}
where $C_{a,b}$ are two integration constants. In order to have a
non-zero $\eta_i$ that obeys the required boundary conditions, we need
to have the following property:
\begin{equation}
\eta_a(-\beta/2,\k^2+g_i)\eta_b(\beta/2,\k^2+g_i)
=
\eta_a(\beta/2,\k^2+g_i)\eta_b(-\beta/2,\k^2+g_i)\; .
\end{equation}
This equation determines the allowed eigenvalues $g_i$. This equation
can also be written as~:
\begin{equation}
\Omega(\beta/2,-\beta/2,\k^2+g_i)=0\; ,
\label{eq:eigen1}
\end{equation}
where $\Omega$ has been introduced in eq.~(\ref{eq:Omega}). The
determinant of the operator $A^*$ is of course obtained as the product
of its eigenvalues:
\begin{equation}
{\rm det}\,A^*[\phi_0,\k^2] = \prod_{g|\Omega(\beta/2,-\beta/2,\k^2+g)=0} g\; .
\end{equation}
(We denote by $A^*[\phi_0,\k^2]$ the restriction of the operator
$A^*[\phi_0]$ to field fluctuations of Fourier mode $\k$.) If we
denote by $z_n$ the (possibly complex) zeros of the function
$\Omega(\beta/2,-\beta/2;z)$, then the solutions of
$\Omega(\beta/2,-\beta/2,\k^2+g)=0$ are the numbers $g=z_n-\k^2$.
Therefore, we can write
\begin{equation}
{\rm det}\,A^*[\phi_0,\k^2] = \prod_n (z_n-\k^2)\; ,
\end{equation}
where multiple zeros are repeated as many times as needed in the
product. The right hand side of this equation is an entire function of
$\k^2$, that obviously vanishes at all the $z_n$'s. Since
$\Omega(\beta/2,-\beta/2,\k^2)$ shares the same property, there exists
an entire function $p(\k^2)$ such that \cite{lang}~:
\begin{equation}
{\rm det}\,A^*[\phi_0,\k^2] = \Omega(\beta/2,-\beta/2,\k^2)\; e^{p(\k^2)}\; .
\end{equation}
From eq.~(\ref{eq:Omega}) we see that, if we chose the functions
$\eta_a$ and $\eta_b$ in such a way that their value at
$\tau=-\beta/2$ is independent of $\k^2$, then the limit
\begin{equation}
\lim_{\k^2 \rightarrow
  \infty}\;|\Omega(\beta/2,-\beta/2,\k^2)|\,e^{-M\sqrt{\k^2}} 
\end{equation}
is bounded for any $M > \beta$. By Hadamard's theorem \cite{lang}, we
conclude that the function $p(\k^2)$ is a constant\footnote{Strictly
speaking, this result only proves the independence of the function $p$
with respect to $\k^2$, but it does not exclude a dependence on the
other parameters of the problem: the mass $m$ and the coupling
constants contained in the potential $U(\phi)$. However, as is clear
from the operator whose determinant we are calculating, this
dependence only arises from the combination $(m^2+U''(\phi_0(\tau))$
which means that it can enter in the final result only via the
solutions $\eta_a$ and $\eta_b$, i.e., via the function $\Omega$. Thus,
the prefactor $\exp(p(\k^2))$ cannot contain any implicit dependence
on these parameters.}. The constant factor $e^p$ must in fact be
proportional to the inverse of the Wronskian of the pair of solutions
$\eta_a$ and $\eta_b$ that we are using, $e^p={\rm const}/\beta W$,
because the determinant must be independent of this choice (the factor
$\beta$ in the denominator has been included by hand in order to have
a dimensionless determinant). The normalization constant can be
absorbed as an overall multiplicative constant in $Z$.

Finally, the determinant of $A^*[\phi_0]$ is obtained by multiplying
the previous result for all $\k$'s, which gives~:
\begin{equation}
{\rm det}\,A^*[\phi_0]
=
\,\exp V\int\frac{d^3\k}{(2\pi)^3}
\ln\left(
\frac{\Omega(\beta/2,-\beta/2,\k^2)}{\beta W}
\right)\; .
\end{equation}
In order to see how the volume $V$ appears in this formula, it is
useful to consider first that the system is in a finite box, and to
rewrite the sum over the corresponding discrete Fourier modes as an
integral.

\section{Integration over the boundary fluctuations}
\label{sec:int_boundary}
\subsection{Expansion of the classical action}
The final step in the analytic part of this calculation is to
calculate the functional integral over the fluctuation $\xi(\x)$ of
the boundary in eq.~(\ref{eq:Z3}). Before doing this integration, we
must expand the classical action $S_{_E}[\phi_c]$ to quadratic order
in $\xi$, using the expansion of eq.~(\ref{eq:phic-exp}) for $\phi_c$.
We have:
\begin{eqnarray}
S_{_E}[\phi_c]=
 S_{_E}[\phi_0]
+
\delta^{(1)}S_{_E}
+
\delta^{(2)}S_{_E}
+{\cal O}(\xi^3)\; .
\label{eq:S-expand}
\end{eqnarray}
Notice that we could be in trouble because a priori we must keep
$\phi_2$ -- the term of order $\xi^2$ in the classical solution
$\phi_c$ -- in the second term of the right hand side, which would be
much more difficult to obtain. We will not need this term however,
because $\phi_0$ is an exact solution of the classical equations of
motion.  Indeed, one can write
\begin{eqnarray}
&&\delta^{(1)}S_{_E}
=
\int (d^4x)_{_E}\;
\frac{1}{2}\left[-\ddot\phi_0(\tau)+m^2\phi_0(\tau)+U^\prime(\phi_0(\tau))
\right](\phi_1(x)+\phi_2(x))
\nonumber\\
&&\qquad\qquad
+\int d^3\x \left[\dot\phi_0(\tau)(\phi_1(x)+\phi_2(x))\right]_{\tau=-\beta/2}^{\tau=+\beta/2}\; .
\end{eqnarray}
The integrand in the first term of the right hand side vanishes
identically because $\phi_0(\tau)$ obeys the classical equation of
motion associated to the action $S_{_E}$. The second term -- a
boundary term -- can be rewritten as follows~:
\begin{equation}
\int d^3\x 
\left[\dot\phi_0(\tau)(\phi_1(x)+\phi_2(x))\right]_{\tau=-\beta/2}^{\tau=+\beta/2}
=
\left[\dot\phi_0(+\beta/2)-\dot\phi_0(-\beta/2)\right]\int d^3\x \xi(\x)\; ,
\end{equation}
and it vanishes because the fluctuation $\xi(\x)$ of the field at the
boundary has a vanishing average. The second order variation of the
action -- the third term in the right hand side of
eq.~(\ref{eq:S-expand}) -- can be written as
\begin{eqnarray}
&&
\delta^{(2)}S_{_E}
=
\frac{1}{2}
\int (d^4x)_{_E}
\Big[\Big(\partial_\mu\phi_1(x)\Big)\Big(\partial^\mu\phi_1(x)\Big)
\!+\!
m^2\phi_1^2(x)
\!+\!
U^{\prime\prime}(\phi_0(\tau))\phi_1^2(x))\Big]
\nonumber\\
&&
\qquad=
\frac{1}{2}
\int (d^4x)_{_E}\;
\partial_\mu\Big[\phi_1(x)\partial^\mu\phi_1(x)\Big]
\nonumber\\
&&
\qquad\qquad
+\frac{1}{2}
\int (d^4x)_{_E}\;
\phi_1(x)\Big[\square_{_E}+m^2+U^{\prime\prime}(\phi_0(\tau))\Big]\phi_1(x)\; .
\end{eqnarray}
The integrand of the second term vanishes because of the equation of
motion obeyed by the field $\phi_1(x)$. Therefore, the second order
variation of the classical action comes entirely from the boundary
term
\begin{eqnarray}
  \delta^{(2)}S_{_E}
  =
  \frac{1}{2}\int d^3\x
  \left[\phi_1(\tau,\x)\partial_\tau\phi_1(\tau,\x)\right]_{\tau=-\beta/2}^{\tau=+\beta/2}\; .
\end{eqnarray}
By rewriting this integral in momentum space, and by making use of the
boundary condition obeyed by $\phi_1(\tau,\x)$ and of
eq.~(\ref{eq:phi1}), we can rewrite this as follows\footnote{In order
to obtain this formula, we use the relation
\begin{equation}
\left[
\partial_{\tau^\prime} G(\tau,\tau^\prime,\k)
\right]_{\tau^\prime=-\beta/2}^{\tau^\prime=+\beta/2}
= \frac{\Omega(\tau,-\frac{\beta}{2},\k^2)
+\Omega(\frac{\beta}{2},\tau,\k^2)}{\Omega(\frac{\beta}{2},-\frac{\beta}{2},\k^2)}
\; .
\nonumber
\end{equation} Therefore, this quantity is equal to $1$ at $\tau=\pm\beta/2$.}~:
\begin{equation}
\delta^{(2)}S_{_E}
  =
\frac{1}{2}\int \frac{d^3\k}{(2\pi)^3} \,C(\k)\;\xi(\k)\xi(-\k)\; ,
\label{eq:dS2}
\end{equation}
where we denote
\begin{equation}
C(\k)\equiv
\Big[\Big[
\partial_\tau\partial_{\tau^\prime}G(\tau,\tau^\prime,\k)
\Big]_{\tau^\prime=-\beta/2}^{\tau^\prime=+\beta/2}
\Big]_{\tau=-\beta/2}^{\tau=+\beta/2}\; .
\label{eq:Ck}
\end{equation}
Therefore, the Gaussian functional integral over $\xi$ leads
to the following result:
\begin{equation}
\frac{e^{-S_{_E}[\phi_0]}\;\sqrt{\beta C({\bf 0})}}{\sqrt{\prod\limits_{\k}\beta C(\k)}}
=
e^{-S_{_E}[\phi_0]}
\;
\sqrt{\beta C({\bf 0})}
\;
\exp\left[-\frac{V}{2}\int\frac{d^3\k}{(2\pi)^3} \ln\left(\beta C(\k)\right)
\right]\; .
\end{equation}
The factor $\sqrt{\beta C({\bf 0})}$ in the numerator serves to remove
the contribution of the zero-modes, i.e., the functions $\xi(\x)$ that
are constant, since these are taken into account in the quantity
$\varphi_0$. Factors of $\beta$ have been introduced in order to make
the arguments of the log and of the square root dimensionless.

\subsection{Calculation of $C(\k)$}
The quantity $C(\k)$ defined in eq.~\eqref{eq:Ck} involves the
calculation of two derivatives of the Green's function evaluated at
the boundaries. This may pose a problem because the derivative of $G$
is not continuous at coincident points. It is crucial to note that
eq.~(\ref{eq:Ck}) imposes a very definite order when taking the limits
$\tau,\tau^\prime\to\pm \beta/2$. This leads to an unambiguous
expression for $C(\k)$~:
\begin{eqnarray}
C(\k)&=&
\lim_{{\tau,\tau^\prime\to+\beta/2}\atop{\tau<\tau^\prime}} 
\partial_\tau\partial_{\tau^\prime}G(\tau,\tau^\prime,\k)
+\lim_{{\tau,\tau^\prime\to-\beta/2}\atop{\tau>\tau^\prime}} 
\partial_\tau\partial_{\tau^\prime}G(\tau,\tau^\prime,\k)
\nonumber\\
&&\quad
-\lim_{{\tau^\prime\to+\beta/2}\atop{\tau\to-\beta/2}} 
\partial_\tau\partial_{\tau^\prime}G(\tau,\tau^\prime,\k)
-\lim_{{\tau^\prime\to-\beta/2}\atop{\tau\to+\beta/2}} 
\partial_\tau\partial_{\tau^\prime}G(\tau,\tau^\prime,\k)\; .
\end{eqnarray}
From eq.~(\ref{eq:G-1}), we see that, depending on the order of $\tau$
and $\tau^\prime$, the double derivative of the propagator reads~:
\begin{eqnarray}
&&
\partial_\tau\partial_{\tau^\prime}G(\tau,\tau^\prime,\k)
=-
\frac{
\partial_{\tau}\Omega(\beta/2,\tau,\k^2)
\partial_{\tau^\prime}\Omega(\tau^\prime,-\beta/2,\k^2)
}{W\;\Omega(\beta/2,-\beta/2,\k^2)}
\qquad\mbox{if\ }\tau^\prime<\tau\; ,
\nonumber\\
&&
\partial_\tau\partial_{\tau^\prime}G(\tau,\tau^\prime,\k)
=-
\frac{
\partial_{\tau^\prime}\Omega(\beta/2,\tau^\prime,\k^2)
\partial_{\tau}      \Omega(\tau,-\beta/2,\k^2)
}{W\;\Omega(\beta/2,-\beta/2,\k^2)}
\qquad\mbox{if\ }\tau^\prime>\tau\; .
\nonumber\\
&&
\end{eqnarray}
Using the explicit form of $\Omega(\tau,\tau^\prime,\k^2)$ given in
eq.~(\ref{eq:Omega}), a straightforward calculation gives~:
\begin{equation}
C(\k)=
\frac{
{\rm det}\,
\begin{pmatrix}
\Delta\eta_a(\k^2) & \Delta\dot\eta_a(\k^2) \cr
\Delta\eta_b(\k^2) & \Delta\dot\eta_b(\k^2) \cr
\end{pmatrix}
}
{
{\rm det}\,
\begin{pmatrix}
\eta_a(\frac{\beta}{2},\k^2) & \eta_a(-\frac{\beta}{2},\k^2) \cr
\eta_b(\frac{\beta}{2},\k^2) & \eta_b(-\frac{\beta}{2},\k^2) \cr
\end{pmatrix}
}\; ,
\label{eq:Ck-1}
\end{equation}
where we denote
\begin{equation}
\Delta\eta_{a,b}(\k^2)\equiv \Big[\eta_{a,b}(\tau,\k^2)\Big]_{\tau=-\beta/2}^{\tau=+\beta/2}\quad,\quad
\Delta\dot\eta_{a,b}(\k^2)\equiv \Big[\dot\eta_{a,b}(\tau,\k^2)\Big]_{\tau=-\beta/2}^{\tau=+\beta/2}\; .
\end{equation}
Notice that the form of $C(\k)$ given in eq.~(\ref{eq:Ck-1}) makes
obvious the fact that $C(\k)$ does not depend upon the choice of the
two solutions $\eta_a$ and $\eta_b$ that one takes, as long as they
are linearly independent. Indeed, the coefficients $C(\k)$ are a
property of the classical action itself, and should be independent on
the basis chosen for the fluctuations around the classical field.

If we take two solutions $\oeta_a$ and $\oeta_b$ such that
\begin{eqnarray}\label{simpleIC}
&&\oeta_a(-\beta/2;\k^2)=1\; ,\quad \dot{\oeta}_a(-\beta/2;\k^2)=0\; ,\nonumber\\
&&\oeta_b(-\beta/2;\k^2)=0\; ,\quad
  \dot{\oeta}_b(-\beta/2;\k^2)=1/\beta\; ,\label{eq:condinit-1}
\end{eqnarray}
then
\begin{equation}\label{eq:C(k)simple}
C(\k) = \frac{\oeta_a(\beta/2)+\beta\,\dot\oeta_b(\beta/2)-2}{\beta\oeta_b(\beta/2)} \,.
\end{equation}
We will suppose that $(m^2 + U'')$ is positive\footnote{In other
terms, the spectrum of the semiclassical propagator has no bound
states \cite{hills-equation}.}.  In this case, one can easily show
from \eqref{eq:ode} that $\oeta_a$ and $\dot\oeta_b$ are monotonically
increasing in $[-\beta/2,\beta/2]$. This implies that $C(\k)>0$ and
$\delta^{(2)}S_{_E} >0$, which means that the fluctuations of the
boundary field always increase the value of the action compared to the
configuration with a uniform boundary condition. This can be seen as
an a posteriori justification for the choice of expanding around
configurations with a uniform boundary condition; indeed, such
configurations have a smaller action than those with fluctuations of
the boundary condition, and thus are the leading contribution to the
partition function.

\subsection{Diagrammatic interpretation}
\begin{figure}[htbp]
\begin{center}
\resizebox*{!}{4cm}{\rotatebox{0}{\includegraphics{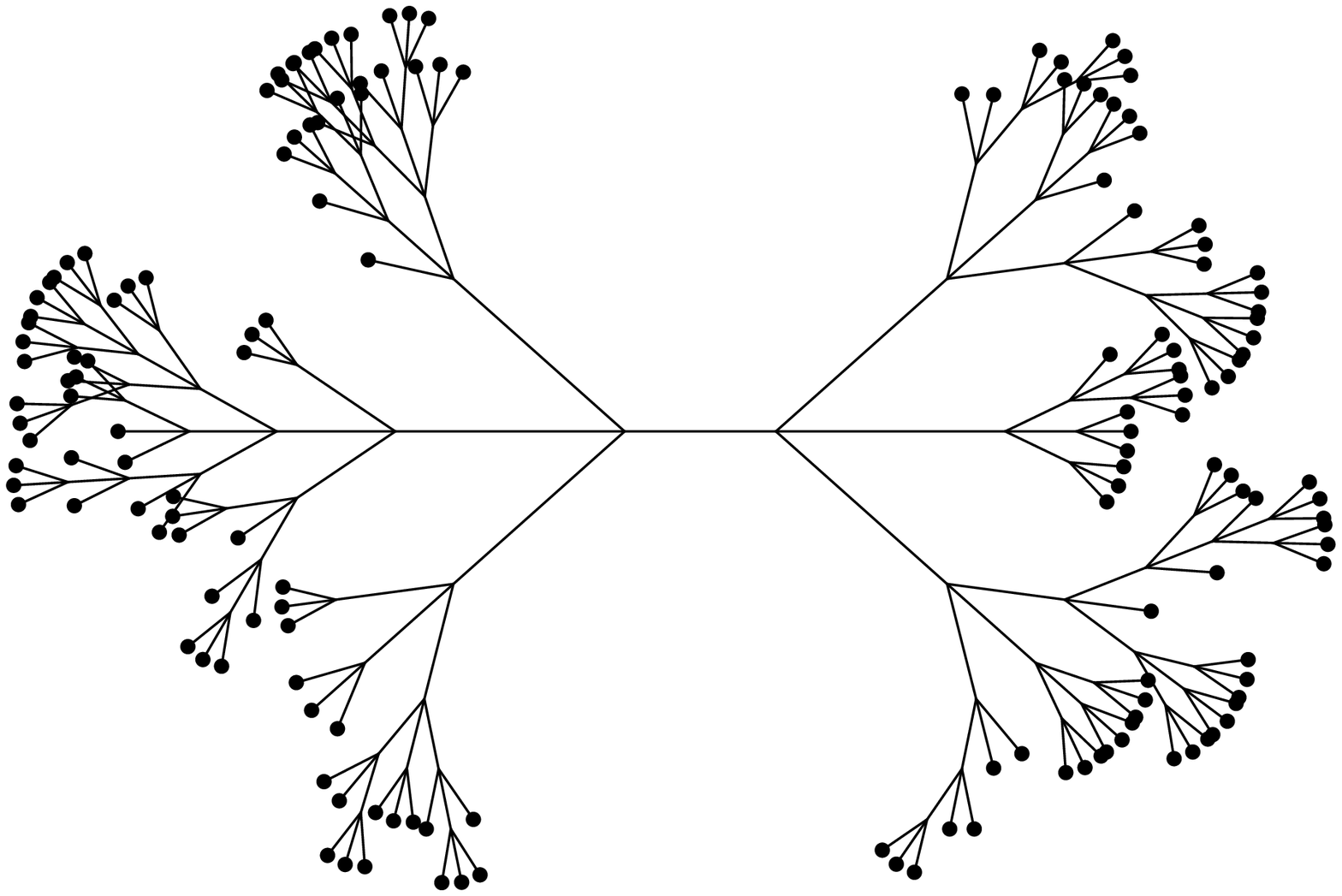}}}
\end{center}
\caption{\label{fig:Se}Diagrammatic expansion of the classical action
  $S_{_E}[\phi_c]$ in terms of the boundary value of the field (black
  dots).}
\end{figure}

The Gaussian integration of $\exp(-S_{_E}[\phi_c])$ over the
fluctuations of the field on the time boundary also corresponds to
some one loop corrections. To begin with, let us recall the obvious
fact that the classical action $S_{_E}[\phi_c]$ only contains terms
that are quadratic or quartic in the classical field $\phi_c$.
Moreover, we have already seen at the end of section
\ref{sec:class_fluct} that the classical field $\phi_c$ is the sum of
all the tree diagrams with one external leg, terminated on the other
side by the boundary field $\varphi$ (see figure
\ref{fig:classfield}).  Thus, $S_{_E}[\phi_c]$ is a sum of tree
diagrams that have no external legs, with the boundary field $\varphi$
at the endpoints of the tree. A typical diagram of that sort has been
represented in figure \ref{fig:Se}.

At this point, these diagrams represent the classical action for an
arbitrary field $\varphi$ as the boundary condition. Writing
$\varphi(\x)=\varphi_0+\xi(\x)$ and doing a Gaussian approximation
means that, for each diagram like the one displayed in figure
\ref{fig:Se}, all the black dots except two of them are replaced by a
uniform boundary field $\varphi_0$ and the remaining two are replaced
by the fluctuation $\xi(\x)$ of the boundary. Then, integrating out
the field $\xi$ means that the endpoints where the $\xi$'s are
attached are linked together, thereby forming a loop. To this loop can
be attached an arbitrary number of tree diagrams terminated by
$\varphi_0$: each of these trees is a contribution to $\phi_0(\tau)$,
the classical solution with boundary value $\varphi_0$. 

Thus, we conclude that the terms resulting from the Gaussian average
over the fluctuations of the boundary field are also 1-loop
contributions in a background made of the field $\phi_0(\tau)$. These
terms are therefore on the same footing as the terms included via the
determinant ${\rm det}\,(A^*[\phi_0])$. Moreover, this analysis of the
diagrammatic content of our approximate expressions confirms the
self-consistency of these approximations: it would have been
inconsistent to keep Gaussian fluctuations of the boundary in ${\rm
  det}\,(A^*[\phi_c])$, because by doing this we would include
two-loop terms in the background field $\phi_0$.

As we shall see in section \ref{sec:renorm}, another consistency check
of our final formula can be made based on the structure of its
ultraviolet divergences: it contains exactly the divergences one
expects of the 1-loop effective action in the background field
$\phi_0(\tau)$, and is thus straightforward to renormalize. It is
important to realize that we need both the 1-loop corrections coming
from ${\rm det}\,(A^*[\phi_c])$, and those coming from the Gaussian
integration over the fluctuations of the boundary field in order to
reproduce the usual pattern of 1-loop ultra-violet divergences.
Failing to include one of the types of terms, one would have spurious
divergences that could not be removed by the usual renormalization
procedure.

\subsection{Final formula for the partition function}
Collecting everything together, we can write the following formula for
the (non-renormalized) partition function:
\begin{eqnarray}
Z&\approx& 
\int\limits_{-\infty}^{+\infty}
d\varphi_0
\; 
e^{-S_{_E}[\phi_0]}\,
\sqrt{\beta C({\bf 0})} \; \nonumber\\
&&\quad\times \;
\exp
-\frac{V}{2}
\int\frac{d^3\k}{(2\pi)^3}
\ln \left[\frac{1}{W}\,\left|
\begin{matrix}
\Delta\eta_a(\k^2) & \Delta\dot\eta_a(\k^2) \cr
\Delta\eta_b(\k^2) & \Delta\dot\eta_b(\k^2) \cr
\end{matrix}
\right|\right]\; .
\nonumber\\
\label{eq:final}
&&
\end{eqnarray}
which is valid for arbitrary choices of $\k^2$-independent initial
conditions. Indeed, the ratio of the determinant and the Wronskian
inside the logarithm does not depend on any particular choice for the
two solutions $\eta_a$ and $\eta_b$. In practice, one can take
advantage of this freedom in order to simplify the numerical
calculations. In particular, for the initial conditions defined in
\eqref{eq:condinit-1} we have\footnote{This can be further simplified
when the potential $U(\phi)$ is an even function of $\phi$. In this
case, we have $\oeta_a(\beta/2,\k^2) =
\beta\dot{\oeta}_b(\beta/2,\k^2)$.}
\begin{eqnarray}
\frac{1}{W}\,
\left|
\begin{matrix}
\Delta\eta_a(\k^2) & \Delta\dot\eta_a(\k^2) \cr
\Delta\eta_b(\k^2) & \Delta\dot\eta_b(\k^2) \cr
\end{matrix}
\right|
\; =\;  
\oeta_a(\beta/2,\k^2) + \beta\,\dot{\oeta}_b(\beta/2,\k^2) -2\;.
\label{eq:simple}
\end{eqnarray}
Thus, we have obtained a fairly compact formula that resums (in the
Gaussian approximation) the fluctuations around the classical solution
and the fluctuations of the boundary condition. At this stage, the
calculation only involves solutions of some ordinary differential
equations, which is in principle straightforward to obtain
numerically.  For each $\varphi_0$, one must determine the following
quantities:
\begin{enumerate}
\item the classical solution $\phi_0(\tau)$,
\item the classical action $S_{_E}[\phi_0]$,
\item for each $\k^2$, two independent solutions $\eta_{a}(\tau,\k)$
and $\eta_{b}(\tau;\k)$ of the equation of fluctuations around the
classical solution $\phi_0(\tau)$.
\end{enumerate}
Notice that all the quantities that depend on $\k$ in fact only depend
on $|\k|$. This means that the integration over $\k$ is in fact a one
dimensional integral.

\section{Renormalization}
\label{sec:renorm}
Our final expression, eq.~(\ref{eq:final}), is plagued by ultraviolet
divergences if taken at face value. These divergences arise from the
integration over the momentum $\k$ in the second line. It is in fact
easy to convince oneself that these divergences can be dealt with by
the usual 1-loop renormalization procedure.  In order to see this, one
must write the solutions $\eta_a$ and $\eta_b$ as series in the
interaction term $U''(\phi_0)$ with the background field. Indeed, if
we denote by $\eta_{a,b}^{(n)}$ the term in $\eta_{a,b}$ that has $n$
powers of $U''(\phi_0)$, we have the following relations~:
\begin{eqnarray}
&&
(\partial_\tau^2-\omega_\k^2)\eta_{a,b}^{(0)}=0\; ,
\nonumber\\
&&
(\partial_\tau^2-\omega_\k^2)\eta_{a,b}^{(n+1)}=U''(\phi_0)\eta_{a,b}^{(n)}
\; .
\end{eqnarray}
{}From these equations, one can see that $\eta_{a,b}^{(n+1)}$ has an
extra power of $1/\k^2$ at large $\k$ compared to
$\eta_{a,b}^{(n)}$. Thus, we expect that only a finite number of terms
in this expansion will actually contain ultraviolet divergences. To
check this, let us calculate explicitly the first three terms in the
expansion of the right hand side of eq.~(\ref{eq:simple}). The
solutions $\oeta_{a,b}^{(0)}$ that obey the boundary conditions of
eq.~(\ref{eq:condinit-1}) are given by~:
\begin{eqnarray}\label{eq:eta_a,b(0)}
&&
\oeta_a^{(0)}(\tau,\k^2)=\cosh\big(\omega_\k(\tau+\frac{\beta}{2})\big)
\; ,\nonumber\\
&&
\oeta_b^{(0)}(\tau,\k^2)=\frac{\sinh\big(\omega_\k(\tau+\frac{\beta}{2})\big)}{\beta\omega\k}\; .
\end{eqnarray}
Notice that these 0th-order solutions already saturate the boundary
conditions at $\tau=-\beta/2$ in eq.~(\ref{eq:condinit-1}). Thus, the
higher order terms in $\oeta_{a,b}$ should vanish and have a
vanishing first time derivative at $\tau=-\beta/2$. In order to find
these terms, it is useful to first construct a Green's function
$\overline{G^0}(\tau,\tau',\k^2)$ of the operator
$\partial_\tau^2-\omega_\k^2$ that obeys the following conditions~:
\begin{eqnarray}
&&
(\partial_\tau^2-\omega_\k^2)\overline{G^0}(\tau,\tau',\k^2)
=\delta(\tau-\tau')
\; ,
\nonumber\\
&&
\overline{G^0}(\tau=-\frac{\beta}{2},\tau^\prime,\k^2)=0\quad,\quad
\partial_\tau \overline{G^0}(\tau=-\frac{\beta}{2},\tau^\prime,\k^2)=0
\; .
\end{eqnarray}
It is straightforward to check that the propagator obeying these
conditions is given by
\begin{equation}
\overline{G^0}(\tau,\tau',\k^2)
=
\theta(\tau-\tau')\frac{\sinh(\omega_\k(\tau-\tau'))}{\omega_\k}\; ,
\end{equation}
which is nothing but the retarded Green's function of
$\partial_\tau^2-\omega_\k^2$. With this Green's function, one can
write
\begin{equation}
\oeta_{a,b}^{(n+1)}(\tau,\k^2)
=
\int_{-\beta/2}^{+\beta/2}d\tau'\;\overline{G^0}(\tau,\tau',\k^2)\;
U''(\phi_0(\tau'))\;\oeta_{a,b}^{(n)}(\tau',\k^2)\; .
\end{equation}
Notice that, since the classical solution $\phi_0(\tau)$ does not depend
on space, the relationship between $\oeta_{a,b}^{(n+1)}$ and
$\oeta_{a,b}^{(n)}$ is local in $\k$.

At this point, it is a straightforward matter of algebra to obtain
$\oeta_{a,b}$ up to second order in $U''$. We obtain
\begin{eqnarray}
&&
\oeta_a(\frac{\beta}{2},\k^2)+\beta\dot\oeta_b(\frac{\beta}{2},\k^2)-2
=
e^{\beta\omega_\k}\Bigg\{
1
\nonumber\\
&&\qquad\qquad
+
\int_{-\beta/2}^{+\beta/2}d\tau'\;\frac{U''(\phi_0(\tau'))}{2\omega_\k}
+
\frac{1}{2}\Bigg[
\int_{-\beta/2}^{+\beta/2}d\tau'\;\frac{U''(\phi_0(\tau'))}{2\omega_\k}
\Bigg]^2
\nonumber\\
&&\qquad\qquad
-
\frac{1}{2}
\int_{-\beta/2}^{+\beta/2}d\tau'd\tau''\;
\frac{e^{-2\omega_\k|\tau'-\tau''|}}{(2\omega_\k)^2}\;
U''(\phi_0(\tau'))\;U''(\phi_0(\tau''))
\nonumber\\
&&\qquad\qquad
+
{\cal O}(e^{-\beta\omega_\k})
+
{\cal O}((U'')^3)
\Bigg\}\; .
\label{eq:uv-expand}
\end{eqnarray}
Inside the curly brackets, we have dropped all the terms that would go
to zero exponentially when $|\k|\to+\infty$. Indeed, these terms do
not contribute to the ultra-violet divergences we are studying in this
section. In this expression, we recognize the time-ordered propagator,
which reads
\begin{equation}
\overline{G_{_{F}}^0}(\tau,\tau',\k^2) =
\frac{e^{-\omega_\k|\tau-\tau'|}}{2\omega_\k}\; .
\end{equation}
It is a remarkable feature of eq.~(\ref{eq:simple}) that, while
having a fairly natural expression in terms of a retarded propagator,
it can be rearranged as an expression involving the time-ordered
propagator (at least for the terms that will contribute to the
ultra-violet divergences).

The terms that appear in the curly bracket in eq.~(\ref{eq:uv-expand})
have a fairly simple interpretation in terms of Feynman diagrams. For
a scalar theory with a $\phi^4$ coupling, the first non-trivial term
can be represented as 
\setbox1\hbox to 1cm{\hfil\resizebox*{0.99cm}{!}{\includegraphics{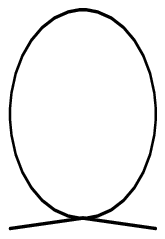}}}
\begin{equation}
\int_{-\beta/2}^{+\beta/2}d\tau'\;\frac{U''(\phi_0(\tau'))}{2\omega_\k}
\quad=\quad\raise -4mm\box1\quad .
\label{eq:tadpole}
\end{equation}
Notice that, in this expression, $1/2\omega_\k$ is the equal-time value
of the time-ordered propagator. Similarly, the term on the third line
can be represented as
\setbox1\hbox to 2.1cm{\hfil\resizebox*{!}{0.99cm}{\includegraphics{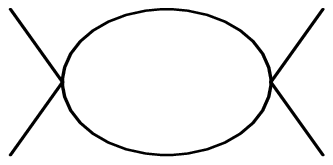}}}
\begin{equation}
-
\frac{1}{2}
\int_{-\beta/2}^{+\beta/2}d\tau'd\tau''\;
\frac{e^{-2\omega_\k|\tau'-\tau''|}}{(2\omega_\k)^2}\;
U''(\phi_0(\tau'))\;U''(\phi_0(\tau''))
\;=\raise -4mm\box1\; .
\label{eq:4point}
\end{equation}
The second term on the second line of eq.~(\ref{eq:uv-expand}) would
be represented by a graph made of two disconnected components, each of
which is given in eq.~(\ref{eq:tadpole}) (the factor $1/2$ is the
symmetry factor that results from the possibility of exchanging the
two connected components). In fact, when we take the logarithm (as
required by eq.~{\ref{eq:final}}), these disconnected contributions
simply drop out~:
\begin{eqnarray}
&&
\frac{1}{2}\ln\Big[
\oeta_a(\frac{\beta}{2},\k^2)+\beta\dot\oeta_b(\frac{\beta}{2},\k^2)-2
\Big]
=
\frac{\beta\omega_\k}{2}
+
\frac{1}{2}
\int_{-\beta/2}^{+\beta/2}d\tau'\;\frac{U''(\phi_0(\tau'))}{2\omega_\k}
\nonumber\\
&&\qquad
-
\frac{1}{4}
\int_{-\beta/2}^{+\beta/2}d\tau'd\tau''\;
\frac{e^{-2\omega_\k|\tau'-\tau''|}}{(2\omega_\k)^2}\;
U''(\phi_0(\tau'))\;U''(\phi_0(\tau''))
+\cdots
\end{eqnarray}
One can check that the cancellation of the disconnected terms when one
takes the logarithm is in fact quite general, and works to all orders.
Finally, when we integrate over $\k$, the first term gives the usual
zero-point energy, and the next two terms are the first two
non-trivial terms of the zero temperature\footnote{We recover the well
known fact that, if a theory is renormalizable at $T=0$, it is also
renormalizable at finite $T$, with the counterterms evaluated at
$T=0$.} 1-loop effective action (for this, it was important to be able
to rewrite the expression in terms of time-ordered propagators). All
these terms are ultra-violet divergent.  If calculated with a momentum
cutoff $\Lambda$, they behave respectively as $\Lambda^4$, $\Lambda^2$,
and $\ln(\Lambda)$, if there are 3 spatial dimensions. All the higher
order terms in the expansion in powers of $U''$ are ultraviolet
finite, because they have at least one extra power of $1/\k^2$ when
$|\k|\to+\infty$.

This identification tells us that, in order to renormalize our final
expression, we must follow the following procedure~:
\begin{enumerate}
\item subtract the ``zero point energy'' in $\ln(Z)$, i.e., subtract
$\beta\omega_\k/2$ from the integrand in the integration over $\k$,
\item add the one-loop counterterms to the classical action
$S_{_E}[\phi_0]$, and simultaneously regularize the integration over
$\k$.
\end{enumerate}
Notice that the regularization scheme employed for calculating the
counterterms must be identical to that used when computing the
integral over $\k$. Thus, a regularization by an ultra-violet cutoff
seems the most convenient method here. Once the above two steps have
been carried out, one will have a $\Lambda$ dependent expression that
tends to a finite result when $\Lambda\to+\infty$.  

This expression of $Z$ is free of any ultra-violet divergence. But,
naturally, it is now expressed in terms of couplings and masses that
are scheme dependent (because one must chose a particular
renormalization scheme\footnote{The renormalization scheme should not
be confused with the regularization scheme.} in order to define
uniquely the counterterms that are added to the classical action). The
standard procedure at this point is to express other physical
quantities in terms of the same scheme-dependent parameters, and to
eliminate them in order to have relationships that involve only
physical quantities.

\section{Conclusions}
\label{sec:conclusions}

We have derived a semiclassical approximation for the partition
function of a system of scalar fields in the presence of an arbitrary
single-well interaction potential. In the path-integral formalism, the
partition function is an integral over periodic configurations in
imaginary time, and is dominated by classical trajectories.  The
non-perturbative information contained in the classical solutions
serves as the starting point for this semiclassical approximation.

Euclidean classical solutions are usually not known for arbitrary
(periodic) boundary conditions.  However, by first considering
classical solutions that correspond to a spatially independent
boundary condition (finding these special solutions amounts to solving
an ordinary differential equation), one can construct approximate
classical solutions obeying arbitrary boundary conditions in a
systematic fashion. We have calculated the contribution of quantum
fluctuations around those classical solutions in a self-consistent
scheme. Our final formula for $Z$ admits a simple expression in terms
of two independent solutions of the equation of small fluctuations around 
the classical solutions, and is thus easily amenable to a
numerical evaluation. Despite its simplicity, our expression treats
exactly the mean value of the field on the boundary, no matter how
large. Moreover, we have shown that this expression is renormalizable
by the subtraction of the standard one-loop counterterms, and by the
subtraction of the free-field energy.

The formula we have obtained for the partition function is
non-perturbative in the sense that it resums the interactions to all
orders for the configurations where the mean value of the field on the
boundary is large. This can be seen by investigating which classes of
diagrams of the usual perturbation theory are taken into account in
our approach. We expect that thermodynamical properties derived from
this semiclassical expression for $Z$ will be valid in a wider domain
in the parameter space $(T,\{\lambda\})$ (where $\{\lambda\}$
represents the coupling constants) as compared to results obtained
from the plain perturbative expansion. We are currently investigating
in detail the case of a theory with a $\lambda \phi^4$
coupling. Results, including a detailed comparison with those obtained
by other resummation schemes, will be presented in a future
publication.

Natural candidates for a direct application of the result
derived in this paper are condensed matter systems containing scalar
order parameters, such as density or magnetization.  Extensions to
potentials with more that one minimum, and other field theories can
also be pursued.

\section*{Acknowledgements}
We would like to thank the financial support of the CAPES-COFECUB
project $443/04$. A.B., C.A.C. and E.S.F. would also like to thank the
support of CAPES, CNPq, FAPERJ and FUJB/UFRJ.

\appendix

\section{The free case}

The action in the free theory is given by
\begin{align}
S[\phi] = \int_{-\beta/2}^{\beta/2}{d^3x d\tau}\bigg [\frac{1}{2}\partial_{\mu}\phi\partial^{\mu}\phi+\frac{1}{2}m^{2}\phi^2\bigg]\;,
\end{align}
leading to the equation of motion:
\begin{align}
\big[\partial^{2}_{\tau} - m^2\big]\phi_{0} =0\;.
\end{align}

The classical solution satisfying $\phi_{0}(-\beta/2)=\phi_{0}(\beta/2)
=\varphi_{0}$ is 
\begin{align}
\phi_{0} = \varphi_{0} \Big[\cosh m(\tau+\beta/2)) + \frac{(1-\cosh(\beta
  m))}{\sinh (\beta m)}\sinh (m(\tau+\beta/2)) \Big]\;.
\end{align}

It is easy to show that $S_{E}[\phi_{0}]=\alpha \varphi_0^2$, with
$\alpha = m V (\cosh(\beta m) -1)$$/$$\sinh(\beta m)$, where $V$ is
the volume. Following our main result, we need two solutions of
\begin{align}
\Big[\partial^{2}_{\tau} - (m^2+k^2)\Big]\eta =0\;,
\end{align}
obeying eq.~(\ref{eq:condinit-1}). We have already seen these
solutions in eq.~(\ref{eq:eta_a,b(0)}).  We obtain
\begin{eqnarray}
\oeta_a(\beta/2,\k) + \beta\,\dot{\oeta}_b(\beta/2,\k) -2 &&= 2(\cosh(\beta\wk)-1)\\\nonumber
&& = (1-\exp(-\beta\wk))^2 \exp(\beta\wk)
\end{eqnarray}
and
\begin{align}
&\beta\,C(\mathbf{0}) = \frac{2m\beta\,(\cosh (\beta m) -1)}{\sinh(\beta m)}\,=\,\frac{2\alpha\beta}{V}\;.
\end{align}

Finally, we have
\begin{align}
Z&\approx \sqrt{\frac{2\alpha\beta}{V}} \int\limits_{-\infty}^{+\infty}
d\varphi_0
\; 
e^{-\alpha \varphi_{0}^{2}}\,
\exp\left [-V\,\int\frac{d^3\k}{(2\pi)^3}\left ( \ln 
 (1-e^{-\beta\wk}) + \frac{\beta\wk}{2}\right )
\right ]\;\\
&=\;\sqrt{\frac{2\pi\beta}{V}}\exp\left [-V\,\int\frac{d^3\k}{(2\pi)^3}\left ( \ln 
 (1-e^{-\beta\wk}) + \frac{\beta\wk}{2}\right ) 
\right ]\;\,
\end{align}
that is (up to an overall constant) the known result for the harmonic
oscillator (not yet renormalized). We see that our approximation
scheme leads to the exact result in the case of the free
theory. Naturally, this is due to the fact that, in the absence of any
interactions, the Gaussian approximation represents exactly the
fluctuations in the system.


\begin{thebibliography}{10}


\bibitem{lebellac}
M. Le Bellac,
{\it Thermal Field Theory}, Cambridge University Press, (2000).

\bibitem{Braaten:2002wi}
   E.~Braaten,
   Nucl.\ Phys.\  A {\bf 702}, 13 (2002).

\bibitem{Kraemmer:2003gd}
   U.~Kraemmer and A.~Rebhan,
   Rept.\ Prog.\ Phys.\  {\bf 67}, 351 (2004).

\bibitem{Andersen:2004fp}
   J.~O.~Andersen and M.~Strickland,
   Annals Phys.\  {\bf 317}, 281 (2005).

\bibitem{Arnold:1994ps}
  P.~Arnold and C.~X.~Zhai,
  Phys.\ Rev.\  D {\bf 50}, 7603 (1994).

\bibitem{Arnold:1994eb}
  P.~Arnold and C.~X.~Zhai,
  Phys.\ Rev.\  D {\bf 51},  1906 (1995).

\bibitem{Parwani:1994zz}
  R.~Parwani and H.~Singh,
  Phys.\ Rev.\  D {\bf 51}, 4518 (1995).

\bibitem{Kajantie:2001hv}
  K.~Kajantie, M.~Laine and Y.~Schroder,
  Phys.\ Rev.\  D {\bf 65}, 045008 (2002).

\bibitem{Braaten:1995cm}
  E.~Braaten and A.~Nieto,
  Phys.\ Rev.\  D {\bf 51}, 6990 (1995).

\bibitem{Braaten:1995ju}
  E.~Braaten and A.~Nieto,
  Phys.\ Rev.\ Lett.\  {\bf 76}, 1417 (1996).

\bibitem{Braaten:1995jr}
  E.~Braaten and A.~Nieto,
  Phys.\ Rev.\  D {\bf 53}, 3421 (1996).

\bibitem{Kajantie:2000iz}
  K.~Kajantie, M.~Laine, K.~Rummukainen and Y.~Schroder,
  Phys.\ Rev.\ Lett.\  {\bf 86}, 10 (2001).

\bibitem{Kajantie:2002wa}
  K.~Kajantie, M.~Laine, K.~Rummukainen and Y.~Schroder,
  Phys.\ Rev.\  D {\bf 67}, 105008 (2003).

\bibitem{Kajantie:2003ax}
  K.~Kajantie, M.~Laine, K.~Rummukainen and Y.~Schroder,
  JHEP {\bf 0304}, 036 (2003).

\bibitem{Karsch:1997gj}
  F.~Karsch, A.~Patkos and P.~Petreczky,
  Phys.\ Lett.\  B {\bf 401}, 69 (1997).

\bibitem{Luttinger:1960ua}
  J.~M.~Luttinger and J.~C.~Ward,
  Phys.\ Rev.\  {\bf 118}, 1417 (1960).

\bibitem{Blaizot:1999ip}
  J.~P.~Blaizot, E.~Iancu and A.~Rebhan,
  Phys.\ Rev.\ Lett.\  {\bf 83}, 2906 (1999).

\bibitem{Blaizot:1999ap}
  J.~P.~Blaizot, E.~Iancu and A.~Rebhan,
  Phys.\ Lett.\  B {\bf 470}, 181 (1999).

\bibitem{Blaizot:2000fc}
  J.~P.~Blaizot, E.~Iancu and A.~Rebhan,
  Phys.\ Rev.\  D {\bf 63}, 065003 (2001).

\bibitem{Blaizot:2003tw}
  J.~P.~Blaizot, E.~Iancu and A.~Rebhan,
  arXiv:hep-ph/0303185.

\bibitem{Peshier:1998rz}
  A.~Peshier, B.~Kampfer, O.~P.~Pavlenko and G.~Soff,
  Europhys.\ Lett.\  {\bf 43}, 381 (1998).

\bibitem{Andersen:1999fw}
  J.~O.~Andersen, E.~Braaten and M.~Strickland,
  Phys.\ Rev.\ Lett.\  {\bf 83}, 2139 (1999).

\bibitem{Andersen:1999sf}
  J.~O.~Andersen, E.~Braaten and M.~Strickland,
  Phys.\ Rev.\  D {\bf 61}, 014017 (2000).

\bibitem{Andersen:1999va}
  J.~O.~Andersen, E.~Braaten and M.~Strickland,
  Phys.\ Rev.\  D {\bf 61}, 074016 (2000).

\bibitem{Andersen:2000yj}
  J.~O.~Andersen, E.~Braaten and M.~Strickland,
  Phys.\ Rev.\  D {\bf 63}, 105008 (2001).

\bibitem{Andersen:2002ey}
  J.~O.~Andersen, E.~Braaten, E.~Petitgirard and M.~Strickland,
  Phys.\ Rev.\  D {\bf 66}, 085016 (2002).

\bibitem{Peshier:1994zf}
  A.~Peshier, B.~Kampfer, O.~P.~Pavlenko and G.~Soff,
  Phys.\ Lett.\  B {\bf 337}, 235 (1994).

\bibitem{Peshier:1995ty}
  A.~Peshier, B.~Kampfer, O.~P.~Pavlenko and G.~Soff,
  Phys.\ Rev.\  D {\bf 54}, 2399 (1996).

\bibitem{rajaraman}
R. Rajaraman,
{\it Solitons and Instantons}, North Holland, (1989).

\bibitem{zinn-justin}
J. Zinn-Justin,
{\it Quantum Field Theory and Critical Phenomena}
(Oxford University Press, 2002).

\bibitem{deCarvalho:1998mv}
   C.~A.~A.~de Carvalho, R.~M.~Cavalcanti, E.~S.~Fraga and S.~E.~Joras,
   Annals Phys.\  {\bf 273}, 146 (1999).
   
\bibitem{deCarvalho:2001vk}
  C.~A.~A.~de Carvalho, R.~M.~Cavalcanti, E.~S.~Fraga and S.~E.~Joras,
  Phys.\ Rev.\  E {\bf 65}, 056112 (2002).

\bibitem{morette1}
C. DeWitt-Morette,
Commun. Math. Phys. {\bf 28}, 47 (1972);
{\bf 37}, 63 (1974); Ann. Phys. (N.Y.) {\bf 97}, 367 (1976).

\bibitem{mizrahi}
M. M. Mizrahi,
J. Math. Phys. {\bf 17}, 566
(1976); {\bf 19}, 298 (1978); {\bf 20}, 844 (1979).

\bibitem{deCarvalho:1999fi}
   C.~A.~A.~de Carvalho, R.~M.~Cavalcanti, E.~S.~Fraga and S.~E.~Joras,
   Phys.\ Rev.\  E {\bf 61}, 6392 (2000).

\bibitem{hills-equation}
W. Magnus and S. Winkler,
{\it Hill's Equation}, Dover, (2004).

\bibitem{lang}
  {S. Lang}, {\sl Complex analysis}, Ed. Springer-Verlag, Berlin (1993).


\bibitem{next}
A. Bessa, C.~A.~A.~de Carvalho, E.~S.~Fraga and F. Gelis,
work in progress.

\end{thebibliography}
\end{document}